\documentclass[
aps, prd, a4paper, 10pt, notitlepage, %reprint,
	amsmath, amssymb, amsfonts, eqsecnum,
	superscriptaddress, 
	nofootinbib, longbibliography % draft bei bedarf
]{revtex4-1}

\usepackage{float}
\usepackage[T1]{fontenc}
\usepackage[utf8]{inputenc}
\usepackage{amsmath}
\usepackage{amsfonts}
\usepackage{amssymb}
\usepackage{subscript}
\usepackage{xcolor}
\usepackage{siunitx}
\usepackage{graphicx}
\usepackage{caption}
\usepackage{textcomp}

\xdefinecolor{mylinkcolor}{rgb}{0 0 0.5}

\usepackage[normalem]{ulem}

\usepackage[pdftex,breaklinks=true,colorlinks=true,filecolor=mylinkcolor,citecolor=mylinkcolor,linkcolor=mylinkcolor,urlcolor=mylinkcolor,menucolor=mylinkcolor,bookmarksnumbered,bookmarksopen,bookmarksopenlevel=1]{hyperref}

\usepackage{tikz}
\usetikzlibrary{snakes}
\usetikzlibrary{shapes,arrows}
\usetikzlibrary{shadows}
\usetikzlibrary{positioning}

\newcommand{\lwtitle}{Precision manufacturing of a lightweight mirror body made by selective laser melting}
\newcommand{\lwshorttitle}{Precision Manufacturing of Lightweight Mirror}
\newcommand{\lwsolid}{0.0}
\newcommand{\nsolid}{Solid}

\newcommand{\lwempty}{68.2}
\newcommand{\nempty}{Empty}

\newcommand{\lwdrill}{25.5}
\newcommand{\ndrill}{Drill}

\newcommand{\lwhexnb}{70.4}
\newcommand{\nhexnb}{HoneyNB}

\newcommand{\lwhexma}{63.5}
\newcommand{\nhexma}{Honey}

\newcommand{\tref}[1]{{table \ref{#1}}}

\newcommand{\Fref}[1]{{Fig. \ref{#1}}}
\newcommand{\fref}[1]{{fig. \ref{#1}}}

\newcommand{\Sref}[1]{{Section \ref{#1}}}
\newcommand{\sref}[1]{{sect. \ref{#1}}}

\begin{document}

\title{\lwtitle}

\author{Enrico~Hilpert}
\email{enrico.hilpert@iof.fraunhofer.de}
\affiliation{\affone}
\affiliation{\afftwo}

\author{Johannes~Hartung}
\affiliation{\affone}

\author{Stefan~Risse}
\affiliation{\affone}

\author{Ramona~Eberhardt}
\affiliation{\affone}

\author{Andreas~T\"unnermann}
\affiliation{\affone}
\affiliation{\afftwo}

\newcommand{\affone}{{Fraunhofer Institute for Applied Optics and Precision Engineering IOF, Albert-Einstein-Stra\ss e 7, 07745 Jena, Germany}}
\newcommand{\afftwo}{{Institute for Applied Physics, Max-Wien-Platz 1, 07743 Jena, Germany}}

\hypersetup{pdftitle={\lwshorttitle},pdfauthor={Enrico Hilpert et al.}}
\date{\today}

\begin{abstract}
This article presents a new and individual way to generate opto-mechanical 
components by Additive Manufacturing, embedded in an established process chain for the
fabrication of metal optics. The freedom of design offered by additive techniques gives
the opportunity to produce more lightweight parts with improved mechanical stability.
The latter is demonstrated by simulations of several models of metal mirrors with
a constant outer shape but varying mass reduction factors. The optimized lightweight
mirror exhibits $\lwhexma\,\%$ of mass reduction and a higher stiffness compared to
conventional designs, but it is not manufacturable by cutting techniques. Utilizing
Selective Laser Melting instead, a demonstrator of the mentioned topological non-trivial
design is manufactured out of AlSi12 alloy powder. It is further shown that -- like in case
of a traditional manufactured mirror substrate -- optical quality can be achieved by diamond
turning, electroless nickel plating, and polishing techniques, which finally results in
$< 150$~nm peak-to-valley shape deviation and a roughness of $< 1$~nm rms in a measurement area of 140$\times$\SI{110}{\micro\metre\squared}. Negative
implications from the additive manufacturing are shown to be negligible. Further it is
shown that surface form is maintained over a two year storage period under ambient
conditions.\vspace{0.5cm}\\

Selective Laser Melting \quad Lightweight Design \quad Diamond Turning \quad Metal Mirror \quad
AlSi12 \quad Additive Manufacturing
\end{abstract}

\maketitle

% INTRODUCTION

\section{Introduction} \label{sec:intro}
Additive Manufacturing (AM) is regarded as the next industrial revolution.
While applications in the medical field exist for some time now, AM has constantly
evolved and is being used in the automotive, civil aviation, military and
aerospace sector today. Due to widespread research the pool of available materials
is constantly growing. Titanium and aluminum alloys are widely used material systems for
structural and lightweight applications due to their low density and high stiffness.
While the machinability of titanium is difficult, aluminum is easy to process
\cite{Machado:Wallbank:1990,Yoder:2006}.
Low density and cost-effective manufacturing render it also a good choice to produce
high performance optical elements.
Traditionally, aluminum mirrors are produced using several manufacturing and finishing
steps resulting in very good shape accuracy (dependent on mirror size) and
roughness values of \SI{<0.5}{\nm} rms \cite{Steinkopf:Gebhardt:Scheiding:others:2008}.
Various aluminum alloys are used for production of the mirror substrates, Al 6061 being the
dominant material due to its high temporal stability \cite{Yoder:2006}.
Targeting applications in the infrared spectral range, precise diamond turned Al
6061 is sufficient as the turning pattern does not interfere with the respective
wavelengths \cite{Yoder:2006,Scheiding:Damm:Holota:others:2010}. For visible and shorter
wavelengths, a polishable layer is necessary to generate smoother surfaces and remove the
turning marks that cause deteriorating scattering of the desired radiation. X-ray
amorphous electroless nickel is a state of the art functional layer for high
performance polishing \cite{Risse:Gebhardt:Damm:others:2008}.
Though, electroless nickel and aluminum alloys differ regarding their
coefficients of thermal expansion. Due to this mismatch shape changes of the part occur
when thermal loads are applied (bimetallic bending).
By using aluminum with \SI{40}{wt\percent} silicon (AlSi40), this thermal mismatch
is reduced to a minimum \cite{Kinast:Hilpert:Lange:others:2014}. 
Beside the optical and mechanical performance of metal mirrors, their weight is an
important factor when highly dynamic scanning applications are desired or when mirrors
are used in optical systems in space. 
The reduction of weight by cutting techniques is a common method but its extent is
limited due to the accessibility of the interior material by manufacturing tools.
Mirrors consisting
of several joined parts or with open backsides are two approaches to increase the mass
reduction at the expense of stiffness.
Yet, monolithic mirrors are ideally suited for space applications due to the absence of
adhesives. Also, a closed backside is desirable as the mechanical stability compared to
open backside mirrors is higher \cite{Ahmad:1996}.
Additive Manufacturing is a promising solution to optimize both mass and mechanical
stability to a higher level than approachable by conventional techniques.
Selective Laser Melting (SLM) is a specific AM technique, which
generates parts out of powder on a layer by layer principle. This enables the 
manufacturing of complex internal structures and thus, offers a high mass saving
potential. The front and back faces of such metal mirrors can remain completely closed,
because there are no cutting tools necessary to generate the lightweight structure. 
In this work, SLM is used to manufacture a hollow structured monolithic metal mirror,
thereby substituting conventional machining processes.
This is realized by employing
aluminum with \SI{12}{wt\percent} silicon (AlSi12) because alongside the also near eutectic
AlSi10Mg alloy it is the most well-understood aluminum material for powder bed based
Additive Manufacturing processes \cite{Buchbinder:Schleifenbaum:Heidrich:others:2011,Louvis:Fox:Sutcliffe:2011,Prashanth:Scudino:Klauss:2014,Vora:Mumtaz:Todd:others:2015,Brandl:Heckenberger:Holzinger:others:2012}.
Besides the generation of the mirror body by the new technique mentioned a 
well-established manufacturing chain for producing a high quality metal mirror is applied.
It should be demonstrated that metal mirrors with optical properties at an ultra-precise 
level are manufacturable by additive processes, also showing the reduction of
mirror mass to values that are not achievable by conventional fabrication techniques.

The article is divided as follows. In \sref{sec:SOA}, a short outline over the state of
the art of AM for optical applications is given. \Sref{sec:designsimeval} contains the
Computer Aided Design (CAD) model descriptions, mandatory definitions necessary for the
investigation, and the evaluation of different mirror design variants per Finite Element
Analysis (FEA). In \sref{sec:manufacturing}, the manufacturing via SLM together with a
quality inspection and the post-finishing are discussed for one single design, also
considering the temporal stability. \Sref{sec:outlook} contains conclusions and an outlook
for future investigations in the context of AM for metal mirrors.\\
All computations are conducted using the finite element program {\scshape ANSYS} \cite{ansys:2016}. 
Further, roughness measurements of machined surfaces are carried out by a Zygo New
View 600 White Light Interferometer, while roughness of the SLM part as-built is
determined by a Taylor Hobson Talysurf profilometer. The shape of the optical surface is
measured using a Zygo GPI XP/D 1000 interferometer. Interior building quality is
analyzed by 3D X-ray tomography utilizing a Phoenix v|tome|x L 240.

% STATE OF THE ART

\section{State of the art} \label{sec:SOA}
Additively manufactured metal mirrors have hardly been researched so far. One study
shows a mirror made of Al 6061, which was replaced by an additive design made
out of Ti6Al4V achieving a mass reduction of $54$~\% using lattice structures
\cite{esa:tropomi:2014}.
In \cite{Sweeney:Acreman:Vettese:others:2015} several approaches are discussed
to manufacture mirrors by additive techniques considering various designs and materials.

Also, the feasibility to use AlSi10Mg and Ti6Al4V to produce mirrors by
additive techniques, including grinding and polishing post-processes, was investigated in
\cite{Herzog:Segal:Smith:others:2015,Mici:Rothenberg:Brisson:others:2015}. 
The main challenges reported by the authors are porosity of the additive manufactured
material as well as the complexity of the CAD models which were developed for the AM
process. Due to the huge amounts of elements, modal and thermal simulations become
complicated to run and require a lot of time.
In spite of that, the potential of tailored designs made by additive manufacturing is
regarded very high. Topology optimized and bionic structures for metal mirrors are also
of particular interest as they offer a good compromise between mechanical functionality
and material usage. These are under current investigation by the authors of
\cite{Herzog:Segal:Smith:others:2015,Mici:Rothenberg:Brisson:others:2015}.\\
In contrast to the design of metal mirrors studies regarding the optimization of
brackets, support structures, and housings for space applications are available in the
literature \cite{Kranz:Herzog:Emmelmann:2014,Rochus:Plesseria:Corbelli:others:2014,Dehoff:Duty:Peter:others:2013}.

% MIRROR DESIGN, SIMULATION and DESIGN EVALUATION

\section{Mirror design, simulation and design evaluation} \label{sec:designsimeval}

% Design study

\subsection{Design study}
\begin{figure}[H]
	\centering
	\includegraphics[width=0.7\textwidth]{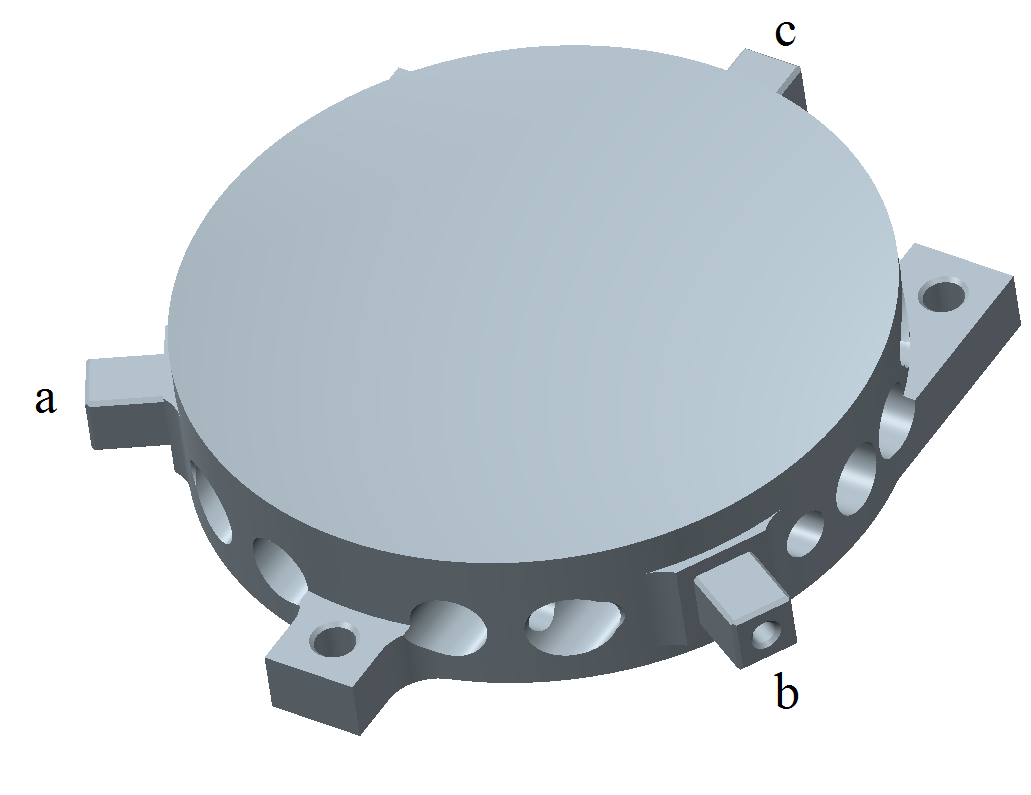}
	\caption{CAD model of the ``drill'' mirror}\label{fig:Kreuz}
\end{figure}

As an initial study, five different mirror designs are investigated. Primarily, the
mirror body is based on a cylindrical geometry, measuring \SI{86}{\mm} in diameter. The
optical surface exhibits a spherical concave shape, with a radius of curvature of
\SI{200}{\mm}, while the backside is flat. Optical front and backside as well as the
circumferential face of the mirror are completely closed. 
The described design represents a full solid mirror without mass reduction.
Additionally, four mass-reduced designs with the same outer dimensions are considered
for a later comparison. The first one is an empty shell model, which represents the
theoretical limit of mass reduction for the present investigation in case of a closed
mirror. Here, the term ``empty'' refers to the complete removal of the interior, while the
remaining shell exhibits wall thicknesses of \SI{2}{\mm} at front and back and \SI{1}{\mm}
at the circumference.
\Fref{fig:Kreuz} shows the second model, which contains holes in a cross directional
pattern along the neutral plane and represents a lightweight design manufacturable by
cutting techniques \cite{Scheiding:Damm:Holota:others:2010,Risse:Gebhardt:Damm:others:2008}.
Third, a novel lightweight design, the ``honeycomb'' mirror, was developed, which is
only manufacturable by Additive Manufacturing. The inner part of the mirror consists of
a hexagon (honeycomb) structure, with additional holes on all faces. Further, this
design features multiple holes with a diameter of \SI{4}{\mm} each on the circumferential
face, which allows a complete removal of any unmolten powder, which remains inside of
the hollow structured mirror body during SLM. Front and backside of the design, as in
the first two models, are completely closed. All relevant wall thicknesses for the honeycomb
design are given in \tref{tab:measures}. The front and back faces are chosen
to be thicker than the circumferential face to ensure a minimum remaining thickness after
manufacturing steps. The thickness of the interior walls is designed to be as thin
as possible, regarding the SLM machine limits.

\begin{table}[H]
	\caption{Measures of the honeycomb mirror in mm, $t$ gives wall thickness, $d\textsubscript{honeycomb}$ represents diameter of inscribed circle of hexagon cells}
	\label{tab:measures}
	\centering
	\begin{tabular}{c|c|c|c|c}
		\centering
		$t\textsubscript{front}$ & $t\textsubscript{back}$ & $t\textsubscript{circumference}$ & $t\textsubscript{honeycomb}$ & $d\textsubscript{honeycomb}$ \\
		\hline
		2.0 & 2.0 & 1.0 & 0.6 & 9.3 \\
	\end{tabular}
\end{table}

\begin{figure}[H]
	\centering
		\begin{minipage}[t]{0.48\textwidth}
			\centering
			\includegraphics[width=1.0\textwidth]{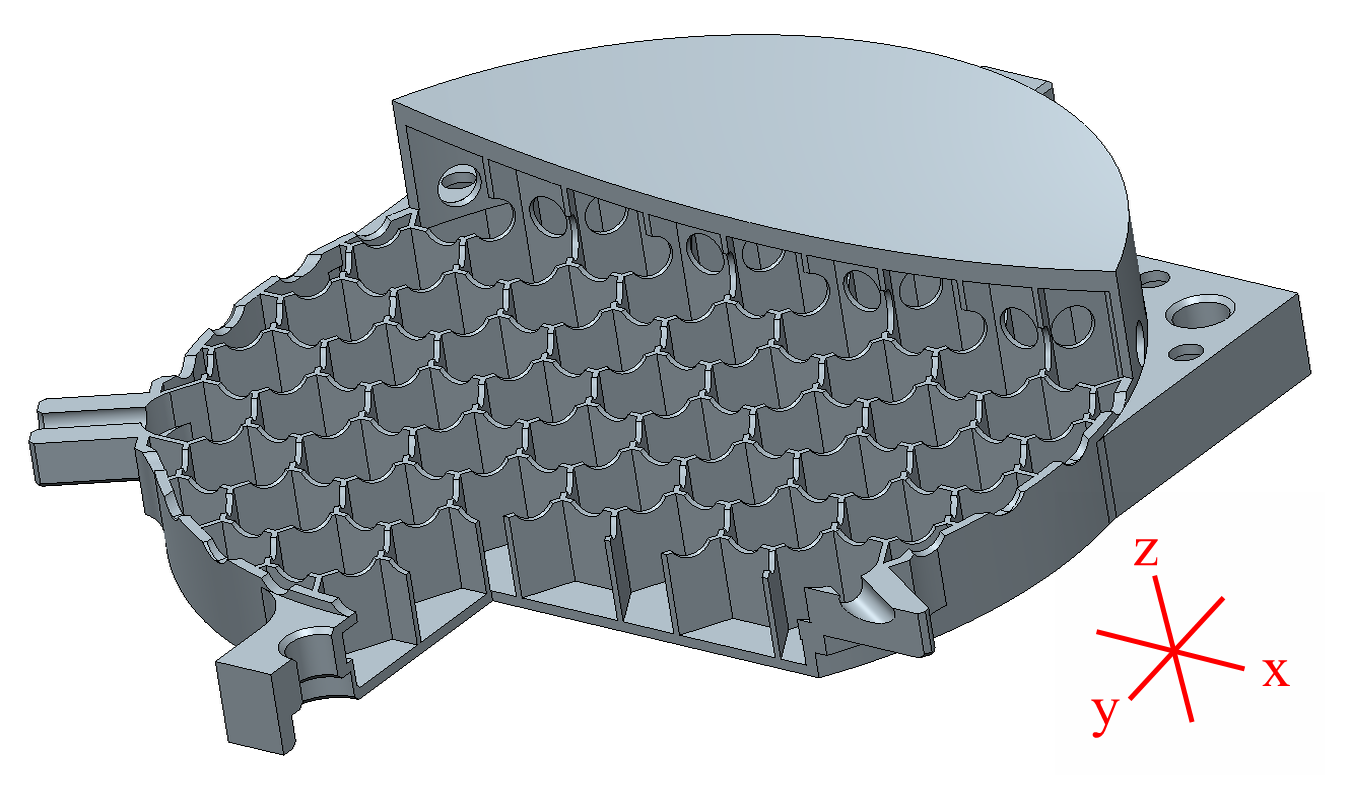}
			\caption{CAD model of the honeycomb mirror with sectioning planes added to demonstrate the hollow structure}
			\label{fig:HexCAD}
		\end{minipage}
		\hfill
		\begin{minipage}[t]{0.48\textwidth}
			\centering
			\includegraphics[width=1.0\textwidth]{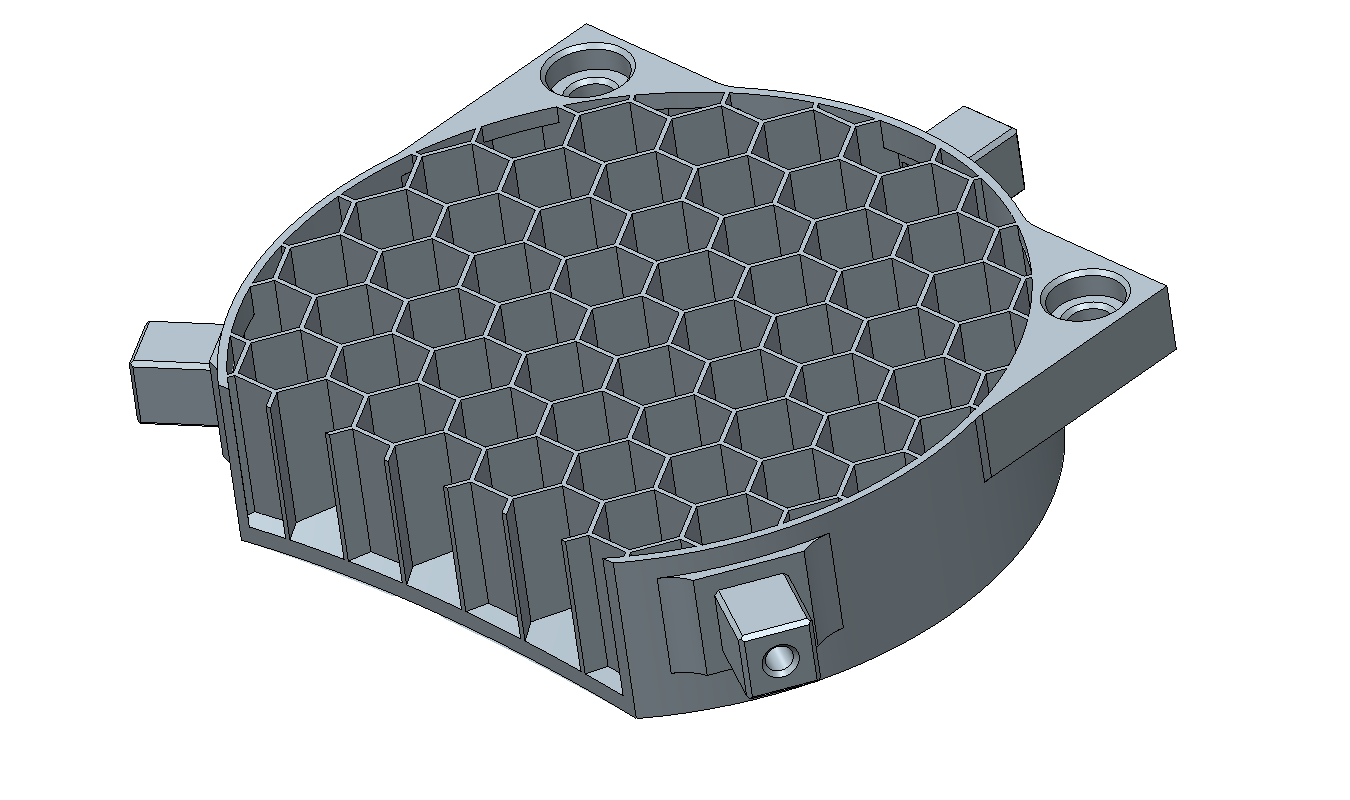}
			\caption{Honeycomb mirror with open backside, upside down}
			\label{fig:HexCADopen}
		\end{minipage}
\end{figure}

The last design is an identical ``honeycomb'' mirror, in this case
with the back side removed, which yields open hexagon cells. This is a common technique for
the production of lightweight mirrors and should therefore serve as a state of the art
design \cite{Ahmad:1996}. This model does not exhibit the inner and circumferential holes of
the previous one as it is manufacturable by conventional cutting techniques.
\Fref{fig:HexCADopen} shows the CAD design from the back.

For evaluating the degree of mass reduction of different designs, the mass reduction factor
$L$ is used, which relates the mass $m$ of the corresponding design variant to the mass
of the solid model $m_\text{\nsolid}$, via
\begin{equation}
	L=1-\frac{m}{m_{\text{\nsolid}}}\label{eq:mass}\,,
\end{equation}
where $L$ is calculated in \%, see \tref{tab:L}.

\begin{table}[H]
	\centering
	\caption{Mass reduction factors [\%] of the mirror designs used in the FEA}
	\label{tab:L}
	\begin{tabular}[c]{c|c|c|c|c}
		\centering
		\nsolid & \ndrill & \nhexma & \nhexnb & \nempty \\
		\hline
		\lwsolid & \lwdrill & \lwhexma & \lwhexnb & \lwempty \\
	\end{tabular}
\end{table}

The ``honeycomb'' mirror (see \fref{fig:HexCAD}) has a mass of about $m_\text{\nhexma}\approx$~\SI{104}{\g}, which corresponds to
$L\approx$~\SI{\lwhexma}{\percent} compared to the solid body
($m_\text{\nsolid}\approx$~\SI{286}{\g}).
All models have three mounting brackets, marked with a--c in \fref{fig:Kreuz}.
These are used as clampings for FEA and as references for
quality inspection, see \sref{sec:manufacturing}.

% Simulation

\subsection{Simulation}
\begin{figure*}
	\centering
	\includegraphics[scale=1.2	]{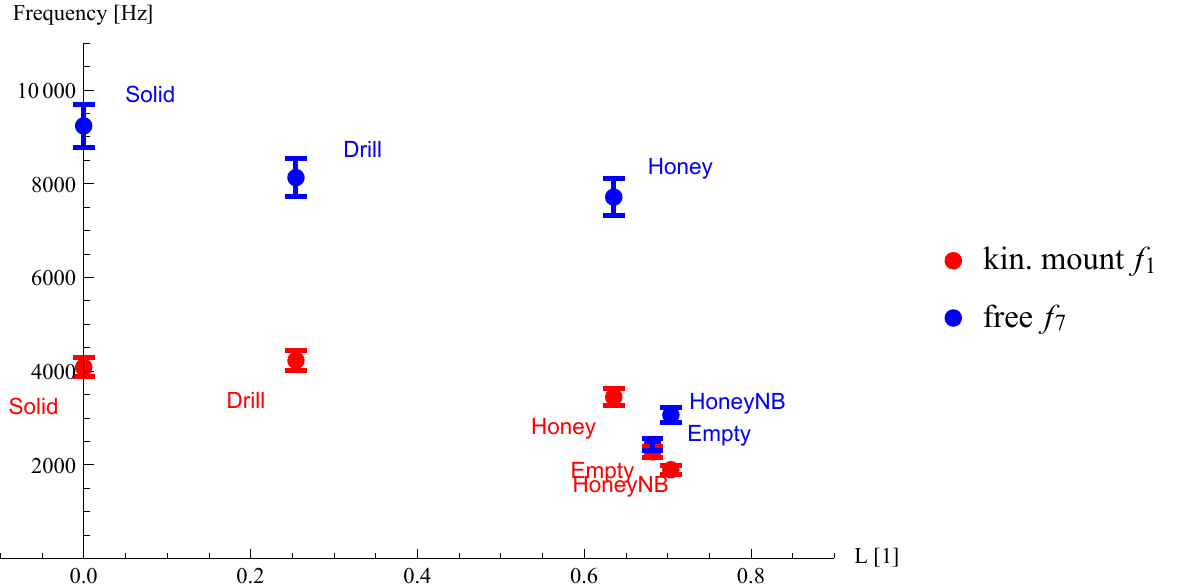}
	\caption{Mass reduction factor versus modal frequency of CAD designs. Errorbars indicate
		deviations from the center eigenfrequency due to variations in Young's modulus $E$ of $\pm$\SI{10}{\percent} which corresponds to $\pm$ \SI{7.5}{\GPa}.}
	\label{fig:Lgraph}
\end{figure*}

To evaluate the stiffness of the mirror designs, a modal analysis is performed using a
slightly simplified model with some curvatures, chamfers, and holes removed.
The mirrors are analyzed in two different mounting setups. These are given
by a kinematical mounting and a setup without any mounting (denoted as ``free''). The
kinematical mounting fixes exactly six degrees of freedom (DOF) of the part, which means
that two tangential DOF are fixed and the radial DOF is unconstrained at the mounting
structures a, b, and c in \fref{fig:Kreuz}. ``Free'' mounting means that translational and
rotational motions may appear under certain load cases. In a modal analysis, this leads to
six zero-modes (one for every DOF).
The resulting eigenfrequencies in connection with the
mass reduction factor $L$ in \eqref{eq:mass} are a measure for the stiffness
of the designs. \Fref{fig:Lgraph} shows the first eigenfrequency $f_1$ and $f_7$, respectively, over $L$ for the
three mirror models in the different mounting setups. The first six zero-modes of the ``free''
setup are omitted (therefore $f_7$).
It is shown that the ``honeycomb'' mirror exhibits a nearly equal stiffness
(eigenfrequency), despite being more lightweight than the ``drill'' model.
The empty shell model and open back structure are distinctly less stable.

% MANUFACTURING AND POST-FINISHING

\section{Manufacturing and post-finishing}\label{sec:manufacturing}
\subsection{Process chain}
After showing the superiority of the ``honeycomb'' design, the manufacturing of the
mirror body is being demonstrated. \Fref{fig:Pchain} shows the process chain for the
production of metal mirrors which will be applied. The conventional machining processes are
complemented by SLM, while the following steps remain feasible on the novel manufactured
mirror body. After additive manufacturing by SLM and stress relieving, the optical surface
of the mirror will be diamond turned, electroless nickel plated, and finished including the
final polishing step.

\tikzstyle{block} = [rectangle, draw, fill=blue!20, 
    text width=0.35\textwidth, text centered, rounded corners, minimum height=3em]
\tikzstyle{line} = [draw, -latex']

\def\deltal{{-2.0cm}}
\def\deltah{{0.5cm}}

% picture definition

\begin{figure*}
\centering
\begin{center}
\resizebox{0.8\textwidth}{!}{
\begin{tikzpicture}[align=center]%, node distance=2.5cm]
	\node[block] (opticaldesign) {{\footnotesize Optical Design}};

	% BLOCK POSITIONS

	\node[block] (mechdesign1) [below left=\deltah  and \deltal  of opticaldesign] {{\footnotesize Classical Mechanical Design}};
	\node[block] (mechdesign2) [below right=\deltah  and \deltal  of opticaldesign] {{\footnotesize Lightweight Mechanical Design}};

	\node[block] (prefab1) [below=\deltah of mechdesign1] {{\footnotesize Conventional cutting fabrication: milling, turning, drilling}};
	\node[block] (prefab2) [below=\deltah of mechdesign2] {{\footnotesize AM (SLM), \\ Selected faces: milling, turning, drilling}};

	\node[block] (spdt1) [below left=\deltah and \deltal of prefab2] {{\footnotesize Diamond Turning}};
	\node[block] (nip) [below=\deltah of spdt1] {{\footnotesize Electroless Nickel Plating}};
	\node[block] (spdt2) [below=\deltah of nip] {{\footnotesize Diamond Turning}};
	\node[block] (mrf) [below=\deltah of spdt2] {{\footnotesize Magnetorheological Finishing}};
	\node[block] (postfinish) [below=\deltah of mrf] {{\footnotesize Chemical Mechanical Polishing}};

	% PATH CONNECTIONS

	\path[line] (opticaldesign) -- (mechdesign1);
	\path[line] (opticaldesign) -- (mechdesign2);

	\path[line] (mechdesign1) -- (prefab1);
	\path[line] (mechdesign2) -- (prefab2);

	\path[line] (prefab1) -- (spdt1);
	\path[line] (prefab2) -- (spdt1);

	\path[line] (spdt1) -- (nip);
	
	\path[line] (nip) -- (spdt2);

	\path[line] (spdt2) -- (mrf);
	
	\path[line] (mrf) -- (postfinish);
\end{tikzpicture}
}
\end{center}

\caption{Process chain to generate mass reduced metal mirrors. After the additive
manufacturing, usually another step is necessary to produce appropriate mounting
planes or other non-optical functional surfaces. Lightweight design in this context
means: the mechanical design cycle for some topological non-trivial interior structures
including an FEA step for optimization.}
\label{fig:Pchain}
\end{figure*}
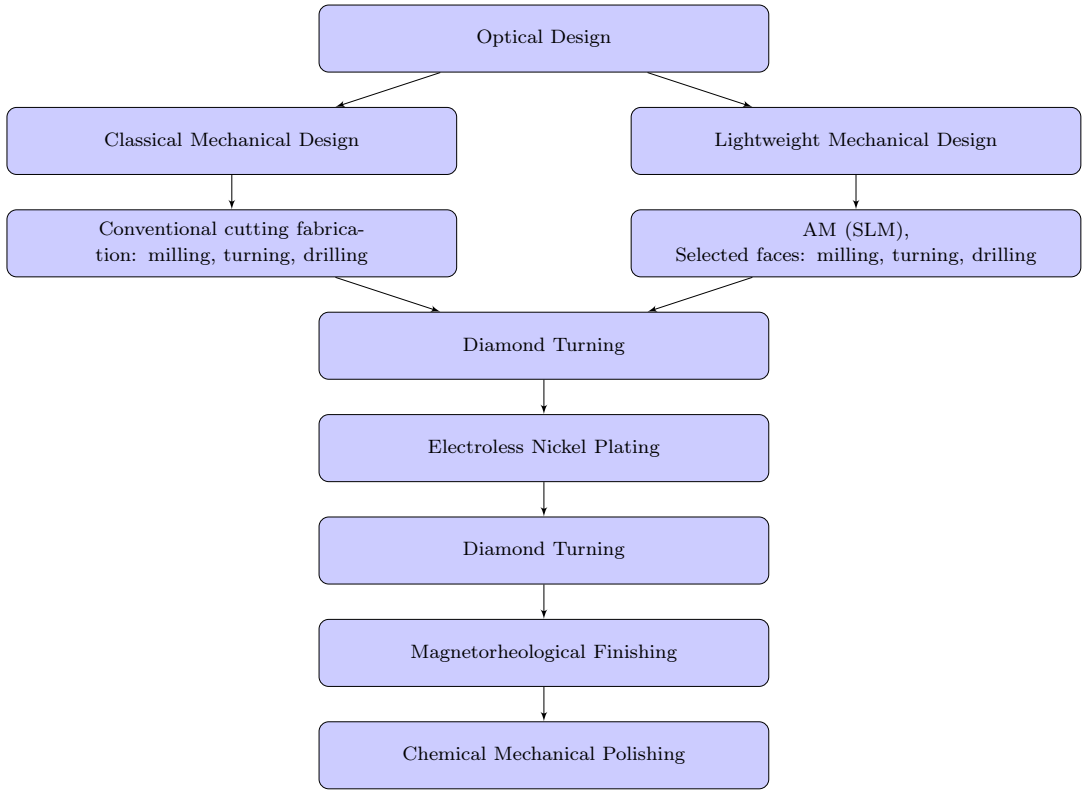

% Additive manufacturing of the mirror body

\subsection{Additive Manufacturing of the mirror body}
\begin{figure}
    \centering
    \begin{minipage}[t]{0.45\linewidth}
        \centering
        \includegraphics[height=6cm]{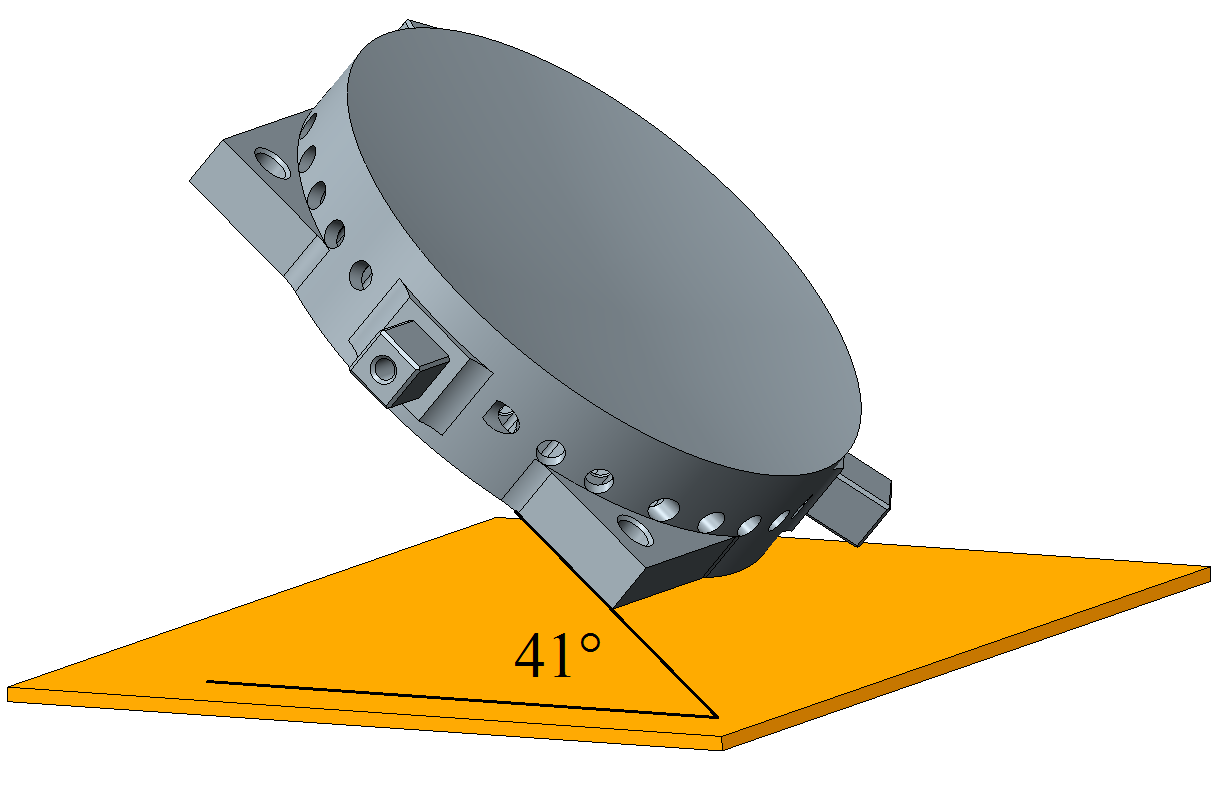}
        \caption{CAD model of honeycomb mirror in building orientation}
        \label{fig:chamber}
    \end{minipage}
    \hfill
    \begin{minipage}[t]{0.45\linewidth}
        \centering
        \includegraphics[height=6cm]{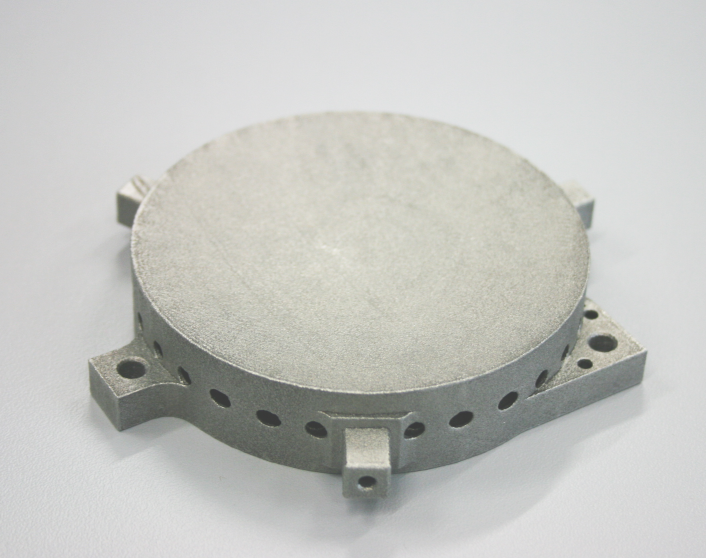}
        \caption{Mirror body after SLM and cleaning}
        \label{fig:SLMmirror}
    \end{minipage}
\end{figure}

In preparation of the manufacturing, the CAD model of the mirror body is virtually aligned
in the building chamber as shown in \fref{fig:chamber}. In order to stabilize the part,
support structures are
used on the outer faces. These are thin lattices of material, which are selective
laser melted along with the mirror during the process and mechanically removed
after finalization. The tilt angle of $41^\circ$ is chosen to minimize overhang features
and critical angles, which are difficult to generate without support structures
(supports have to be avoided in the mirror interior) \cite{Kruth:Mercelis:VanVaerenbergh:others:2007,Su:Yang:Xiao:2012}. 
The term ``overhang'' refers to all faces whose surface normals are pointing towards the
building platform (orange box in \fref{fig:chamber}). A Concept Laser M2 Cusing SLM machine
is used to build the mirror
substrate. AlSi12 alloy powder with a particle size of \SI{<25}{\micro\metre}
and spherical shaped particles is applied for manufacturing, utilizing nitrogen as
shielding gas. The layer thickness measures \SI{25}{\micro\metre} for the actual part and
\SI{50}{\micro\metre} for the support
structures, which means supports are generated every second layer. The illumination of the
part is done in a continuous pattern, which means single lines are scanned next to each
other in the volume, finalized by a contour line on the outer and inner surfaces. Details
about the laser parameters (scanning velocity, average power, hatching distance) are not
available as they are protected by the machine vendor.\\
After SLM the powder is removed from the interior of the mirror
using the holes on the circumferential and interior faces. First, this is carried out while
still working in the glovebox of the SLM machine under inert atmosphere. Next step is the
separation of the mirror from the building platform and support structures using a saw. Due
to the utilization of tapered geometry at the joint to the mirror, the remaining support
structures could be easily separated manually. Eventually, cleaning procedures could
be carried out, which included several wet rinsing and ultrasonic steps.
\Fref{fig:SLMmirror} shows the mirror substrate at this point. Considering the cleaning it
shall be noted that the size of the holes should be increased in order to promote better
fluid flow and reduce cleaning time.

% Quality inspection

\subsection{Quality inspection}
The outer surface of the mirror body shows
a typical roughness for SLM processed parts (\SI{>5}{\micro\metre} $R_\text{a}$) and strongly depends on
the orientation of structures with respect to the powder layers or building direction. In
general, all faces that are oriented towards the building platform show increased
roughness, which can be mainly attributed to powder particles adhering to the melt
pool from the bottom. Similar results were found by other researchers, too
\cite{Kruth:Mercelis:VanVaerenbergh:others:2007}. 
The outer holes on the circumferential face show elliptical shape with blunt edges, which
is a result of overhanging geometry. The measures show variations of up to
\SI{200}{\micro\metre} in positive direction mostly (excess material). As a consequence
outer bores are too small and show high roughness on overhanging areas.
This, under some circumstances, makes post manufacturing necessary, which arises the question
whether outer bores should be generated by SLM at all. This should be decided depending on the tolerances needed.

\begin{figure}[H]
	\centering
	\includegraphics[width=0.6\textwidth]{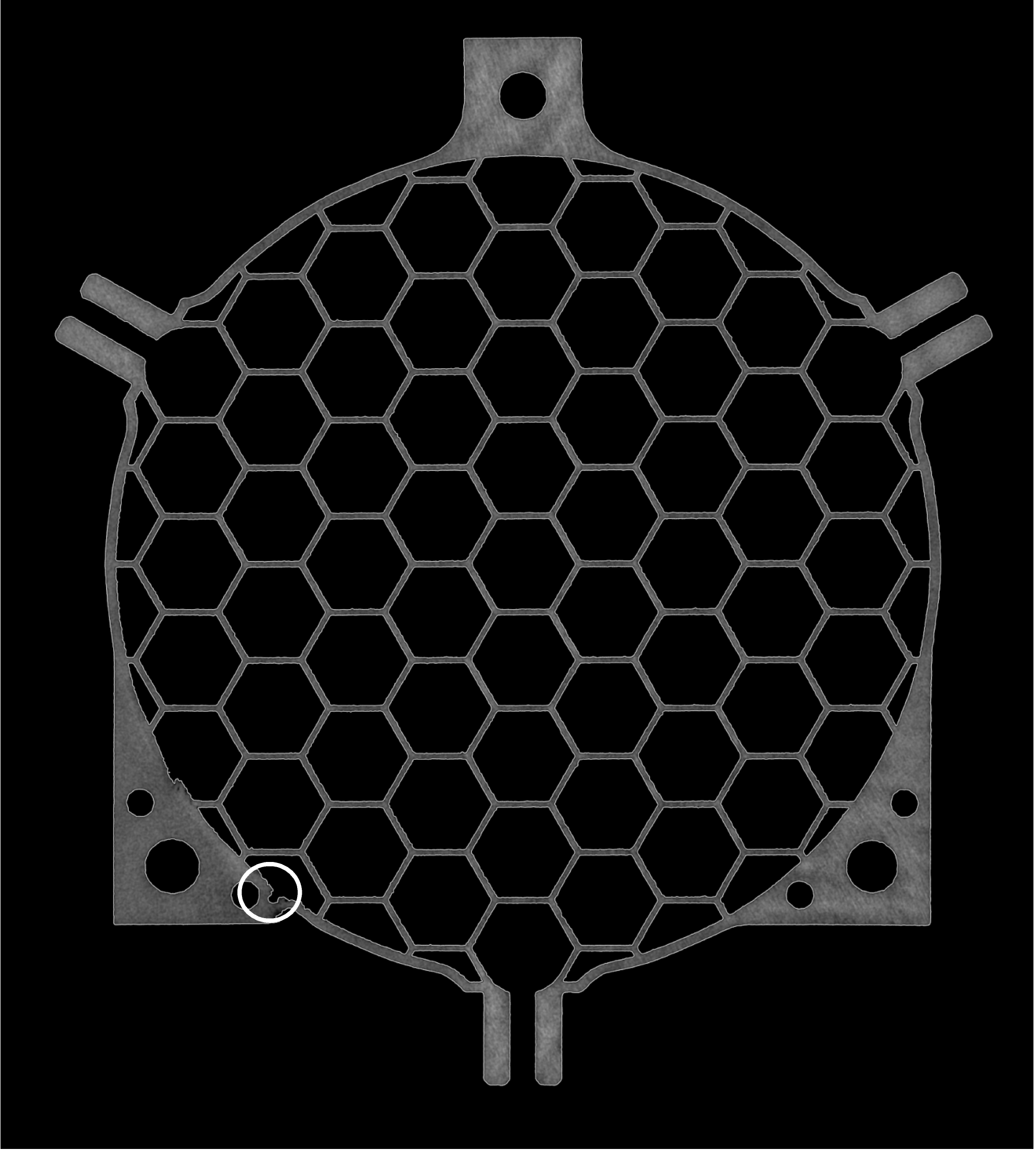}
		\caption{Building error revealed by X-Ray tomography}
	\label{fig:Pore}
\end{figure}

Internal geometry is examined using 3D computed tomography \cite{Wiemuth:2013}. This
generates a three dimensional voxel model, which is then compared to the CAD model. The
size of the scanned part allows a scanning resolution of \SI{60}{\micro\metre}. The scan is
performed in the $xy$ and the $yz$ plane, with reference to the coordinate system in
\fref{fig:HexCAD}. The data shows that the inner hollow structure is generated completely.
Coarse errors such as porosity, missing or unintended structures are analyzed
by evaluating the two-dimensional cross sectional images. Only one pore (see
\fref{fig:Pore}, white circle) is detected in the images, which shows a good building
quality. Its size of \SI{1.5}{\mm} is determined by analyzing the X-ray
pictures using the voxel size. 
Due to the extent of the pore, its origin is most likely caused by an erroneous layer
generation during SLM, for example because of missing powder at
this specific location. The powder coating is done by a rubber blade, which is prone to
warping of already solidified material. If the part warps (because of residual stress) the
rubber bends around it and quickly moves onward once the warped area is passed. This leads
to zones of missing powder directly behind this geometry. Also, material evaporation
by the laser radiation can be excluded as an origin of the pore because this type of pores
cannot be larger than the melt pool. An online monitoring of the SLM process (e.g. camera
images of each layer) could provide more details about such defects, but was not used during
the building. 
Other inhomogeneities such as small grooves or increased roughness (adherence of particles)
are occasionally found in the X-ray images.

% Finishing operations

\subsection{Finishing operations}
\begin{figure}
	\centering
		\includegraphics[width=0.6\textwidth]{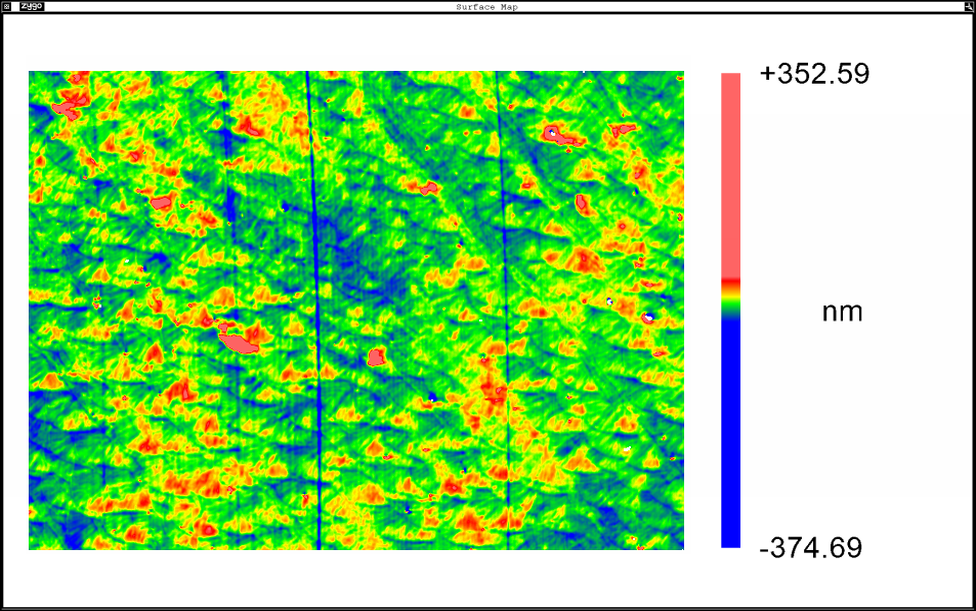}\\
			\begin{tabular}{ccc}
				$Sz$ [nm] & $Sq$ [nm] & $Sa$ [nm]\\
				727.3 & 12.7 & 9.1 \\
			\end{tabular}
		\caption{Surface roughness @ 2.8$\times$\SI{2.1}{\mm\squared} after diamond turning of AlSi12 body; feed direction: horizontal, tangential direction: vertical}
		\label{fig:RALU}
\end{figure}

Subsequent to the SLM process, the cleaned mirror substrate is artificially aged using an
appropriate heat treatment. This is necessary to reduce stress induced during the laser
melting process, which is caused by very high thermal gradients. These effects are an
inherent problem of many additive manufacturing processes and therefore subject to
intensive research \cite{Vora:Mumtaz:Todd:others:2015,Buchbinder:Schilling:Meiners:others:2011,Vrancken:Cain:Knutsen:others:2014,Mercelis:Kruth:2006,Bo:Yu-sheng:Qing-song:others:2012}. After heat treatment the front and back sides of the mirror have to be milled.
The part, showing near net shape already, is therefore aligned using the circumferential face for centering and the mounting structures (a--c, see \fref{fig:Kreuz}) to get the $z$ position. In this step support structures from the back face are removed
and the front surface is smoothened for diamond turning. The closed backside of the mirror
substrate renders advantageous for this technique, because a vacuum chuck is applicable for
mounting on the turning machine. For proper function, planarity of the backside is
established using a lapping process after the milling. By utilizing this procedure,
a mechanical mounting or application of adhesives is avoided.
The following ultra-precise diamond turning is carried out in two passes. This reduces the
roughness of the optical surface and establishes the target shape, which allows the
measurement of both parameters by optical techniques.
The resulting surface roughness is measured using
a white light interferometer with a 2.5$\times$ lens covering an area of
2.8$\times$\SI{2.1}{\mm\squared}, see \fref{fig:RALU}. Turning marks are hardly visible
(feed direction is horizontal, tangential direction is vertical in the image), which can be
attributed to brittle silicon particles leading to high cutting forces and therefore
rougher surfaces compared to Al 6061 \cite{Steinkopf:Gebhardt:Scheiding:others:2008,Ahmad:1996}. As a result, the value of $Sz$ is very large compared to the averaged roughness
values, which can be attributed to outliers due to the silicon particles.

\begin{figure}
	\centering
		\includegraphics[width=0.6\textwidth]{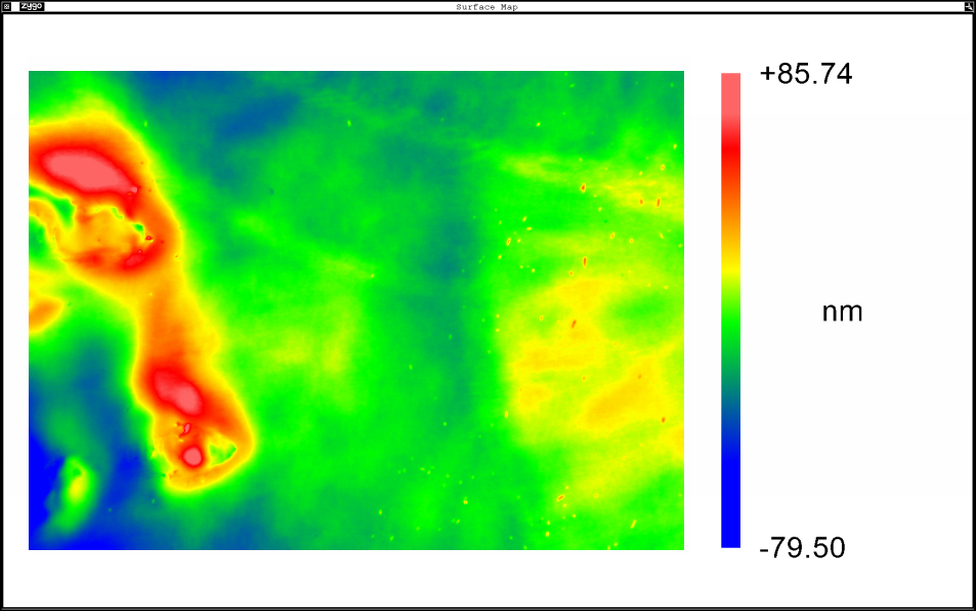}\\
			\begin{tabular}{ccc}
				$Sz$ [nm] & $Sq$ [nm] & $Sa$ [nm]\\
				165.2 & 19.8 & 14.7 \\
			\end{tabular}
		\caption{Surface roughness @ 140$\times$\SI{110}{\micro\metre\squared} after diamond turning of AlSi12 body}
		\label{fig:RALU2}
\end{figure}

\Fref{fig:RALU2} shows the roughness using a 50$\times$ lens, covering an area of
140$\times$\SI{110}{\micro\metre\squared}. An elevated particle is visible, which mainly contributes to the roughness.

The polycrystalline and multi-phase AlSi12 material can not be manufactured to
sufficient roughness values. Therefore, an electroless nickel layer is plated onto
the complete mirror substrate to surmount these limitations. The process parameters
during electroless nickel plating have been tailored to incorporate 11 to \SI{12}{wt\percent} phosphorus into the
layer. As a result, the material is X-ray amorphous \cite{Kanani:2007}. This condition
in combination with resulting hardness values of 500 to 600~HV~0.1 ensures good
machinability of the mirror surface \cite{Mallory:Hajdu:1990,Taylor:Syn:Saito:others:1986,Dini:1992}.
The plating duration was adjusted to achieve a thickness of \SI{100}{\micro\metre}, which provides a dense
structure even after removal of several micrometers of material by the following
manufacturing steps \cite{Kanani:2007,Walsh:Leon:Kerr:others:2008}. 

\begin{figure}
	\centering
	\includegraphics[width=0.6\textwidth]{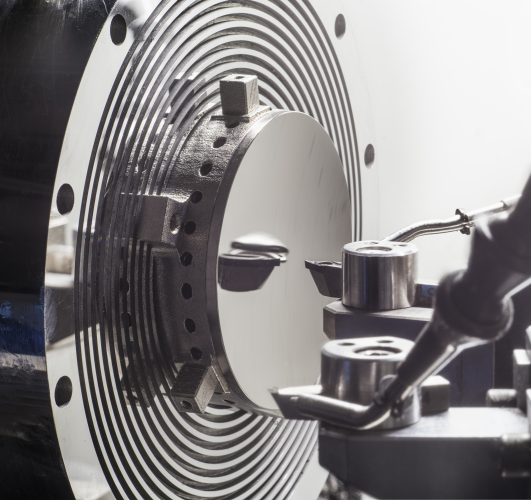}
	\caption{Mirror mounted on diamond turning machine}
	\label{fig:final}
\end{figure}

\begin{figure}
	\centering
	\includegraphics[width=0.6\textwidth]{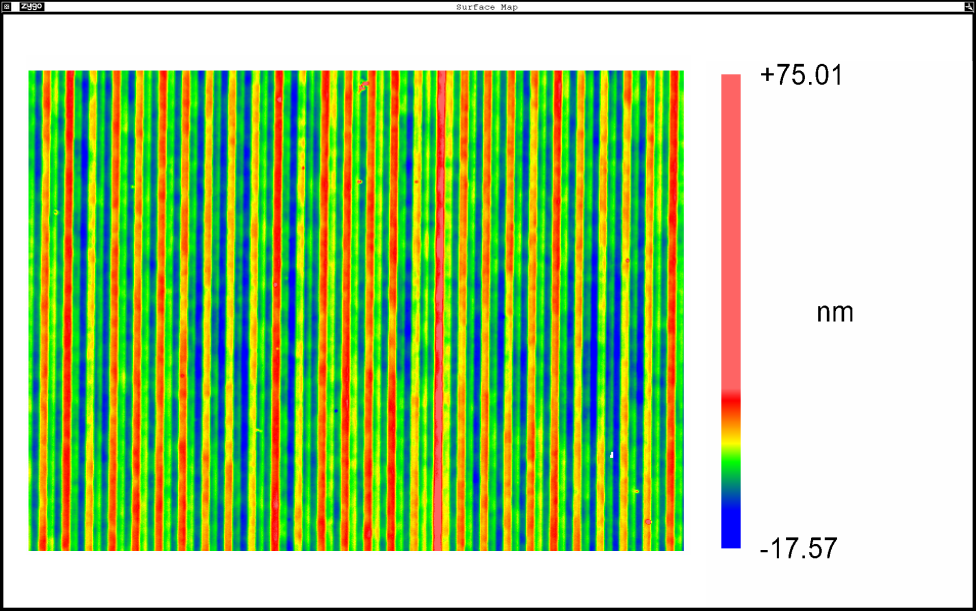}\\
		\begin{tabular}{ccc}
				$Sz$ [nm] & $Sq$ [nm] & $Sa$ [nm]\\
				92.6 & 5.4 & 4.5 \\
		\end{tabular}
	\caption{Surface roughness @ 140$\times$\SI{110}{\micro\metre\squared} after diamond turning of electroless nickel plated mirror substrate}
	\label{fig:NiP-UP}
\end{figure}

After the plating procedure, the mirror is heat treated and undergoes temperature cycling.
A subsequent diamond turning procedure reestablishes the target shape and reduces
roughness after plating. \Fref{fig:final} shows the mirror mounted on the precision
diamond turning machine at this step. The roughness after turning measures \SI{5.4}{\nm}
$Sq$ in the 140$\times$\SI{110}{\micro\metre\squared} area, the turning marks are well
defined, see \fref{fig:NiP-UP}.

\begin{figure}
		\begin{minipage}[t]{0.45\textwidth}
			\includegraphics[height=5.0cm]{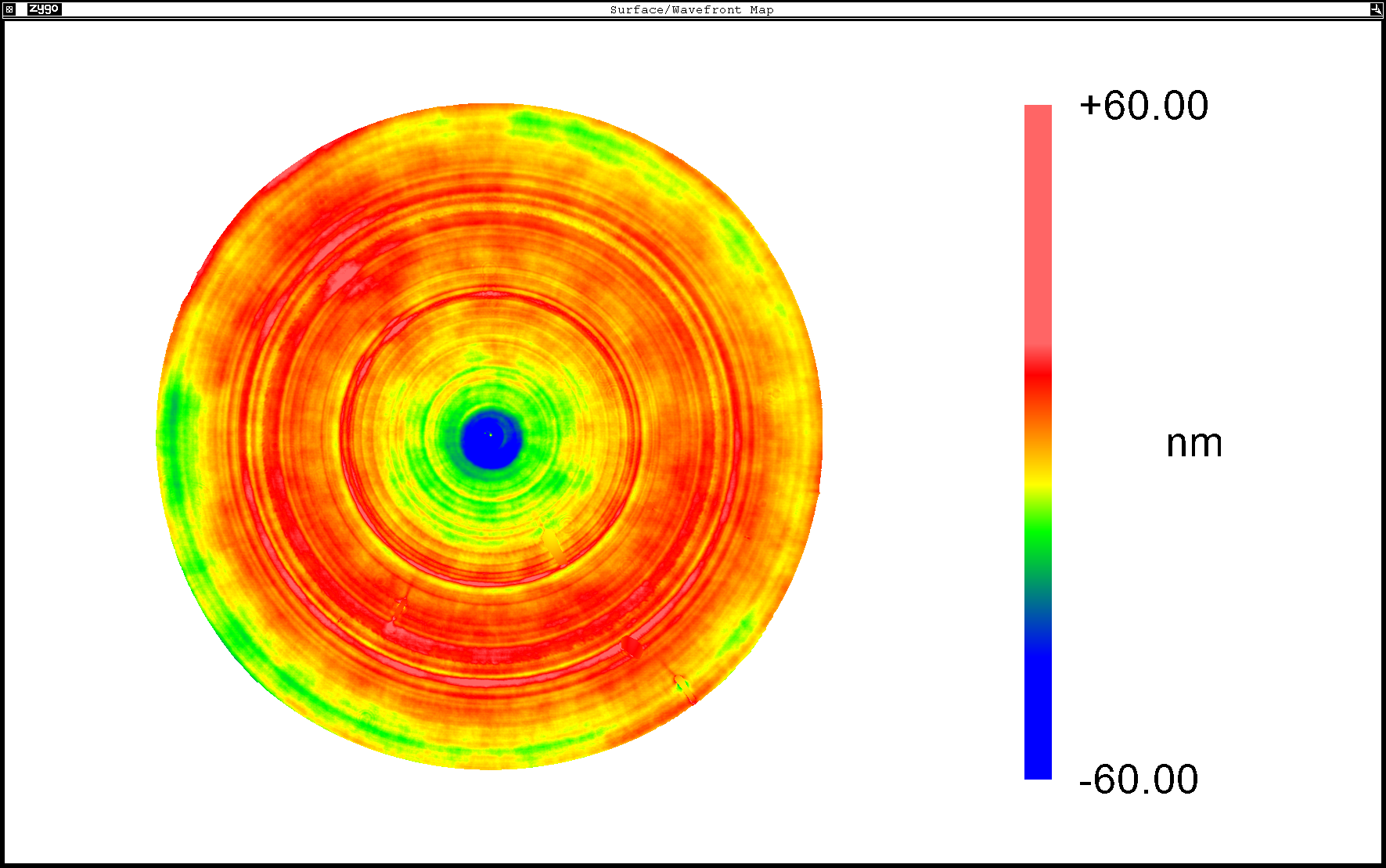}
				\begin{tabular}{cc}
					$p.-v.$ [nm] & $rms$ [nm] \\
					110.4 & 8.2 \\
				\end{tabular}
			\caption{Surface shape deviation after MRF}
			\label{fig:FormMRF}
		\end{minipage}
		\hfill
		\begin{minipage}[t]{0.45\textwidth}
			\includegraphics[height=5.0cm]{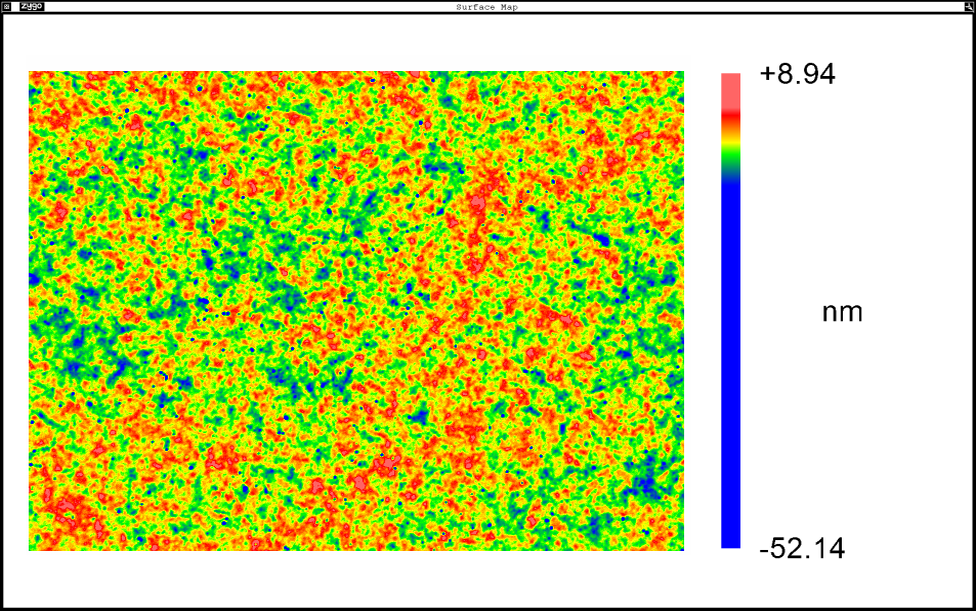}
				\begin{tabular}{ccc}
					$Sz$ [nm] & $Sq$ [nm] & $Sa$ [nm] \\
					61.1 & 1.9 & 1.4 \\
				\end{tabular}
			\caption{Surface roughness @ 140$\times$\SI{110}{\micro\metre\squared} after MRF}
			\label{fig:RMRF}
		\end{minipage}
\end{figure}

\begin{figure}
		%\centering
		\begin{minipage}[t]{0.45\textwidth}
			\includegraphics[height=5.0cm]{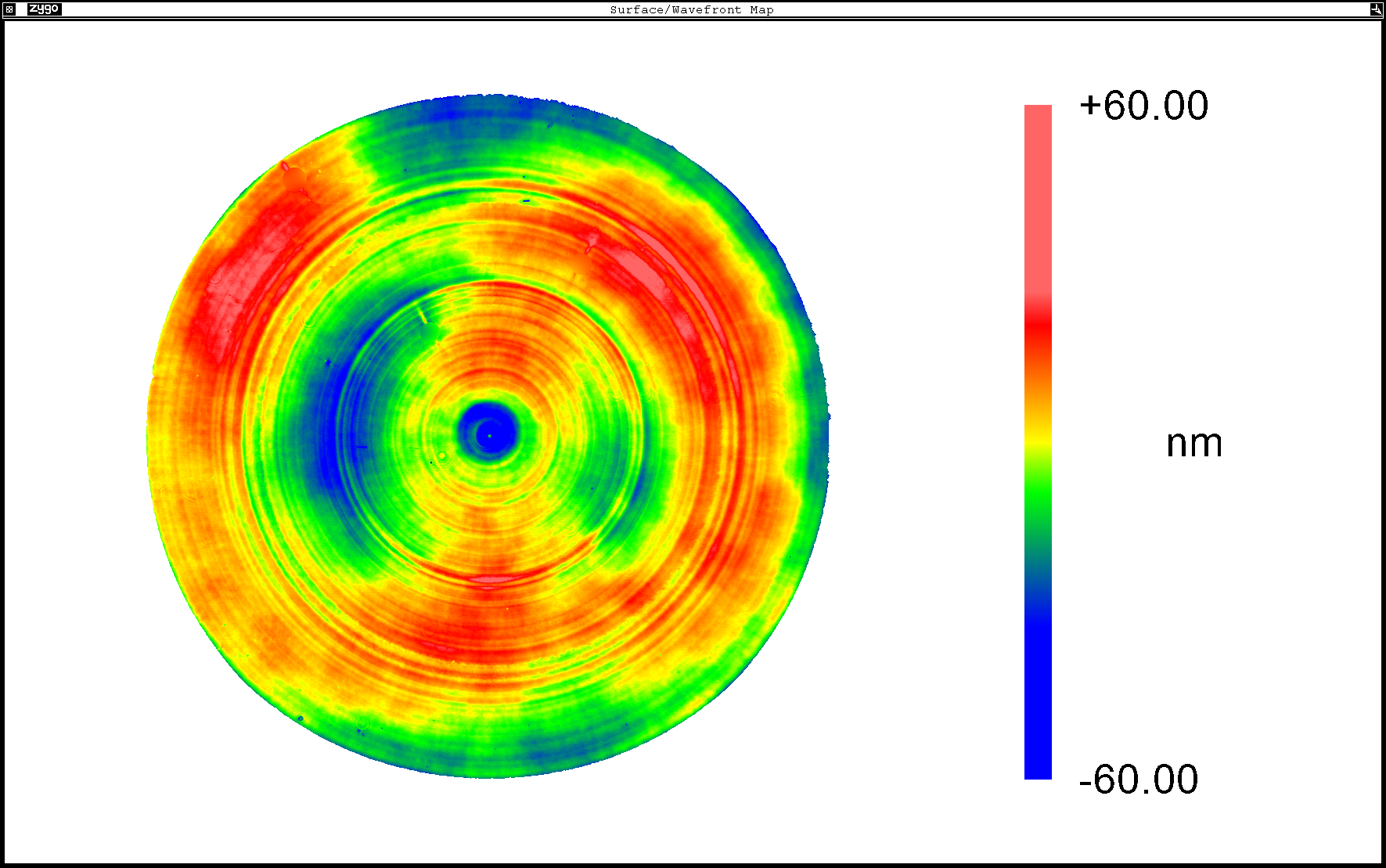}
				%\centering
					\begin{tabular}{cc}
						$p.-v.$ [nm] & $rms$ [nm] \\
						108.9 & 12.5 \\
					\end{tabular}
			\caption{Surface shape deviation after CMP}
			\label{fig:FormCMP}	
		\end{minipage}
		\hfill
		\begin{minipage}[t]{0.45\textwidth}
			\includegraphics[height=5.0cm]{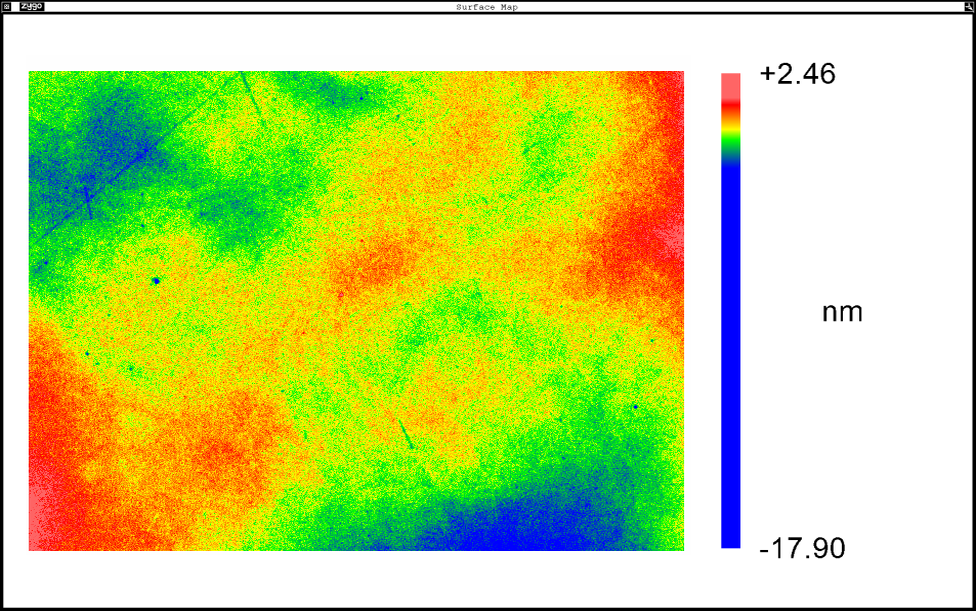}
				%\centering
				\begin{tabular}{ccc}
					$Sz$ [nm] & $Sq$ [nm] & $Sa$ [nm]\\
					20.4 & 0.6 & 0.4 \\
				\end{tabular}
			\caption{Surface roughness @ 140$\times$\SI{110}{\micro\metre\squared} after CMP}
			\label{fig:RCMP}
		\end{minipage}
\end{figure}

In a next step the optical surface is processed by Magnetorheological Finishing (MRF),
which is capable of performing local shape corrections with high accuracy \cite{Beier:Scheiding:Gebhardt:others:2013}. The mechanism of local shape correction on
electroless nickel plated metal mirrors as well as further process details can be found in
\cite{Beier:Fuhlrott:Hartung:others:2016}.
\Fref{fig:FormMRF} shows the interferometric measurement of the shape after this step. 
For analysis of the shape deviation, alignment contributions piston, tip/tilt, and power are subtracted from the resulting data. The remaining error 
of \SI{110.4}{\nm} peak-to-valley and \SI{8.2}{\nm} rms therefore represents the deviation from the best-fit spherical shape.
In order to remove 
edge effects from manufacturing and measurements, the clear aperture was set to \SI{81}{\mm} for all shape analyses.
The surface roughness @ 140$\times$\SI{110}{\micro\metre\squared} is reduced to \SI{1.9}{\nm} $Sq$, showing an island-type
morphology, which could be an indication of local variations in material removal rate, see
\fref{fig:RMRF}. After correction of the shape and improvement of roughness of the
optical surface, Chemical Mechanical Polishing (CMP) is carried out to smoothen the optical
surface, which further reduces the roughness. \Fref{fig:FormCMP} shows the shape
deviation after polishing, which is only slightly influenced compared to the state after
MRF. The roughness is improved to values of \SI{<0.6}{\nm} $Sq$, see \fref{fig:RCMP}.

% temporal stability
\subsection{Final evaluation and temporal stability}
\begin{figure}
	%\centering
	\begin{minipage}{0.6\textwidth}
		%\centering
		\includegraphics[width=1.0\textwidth]{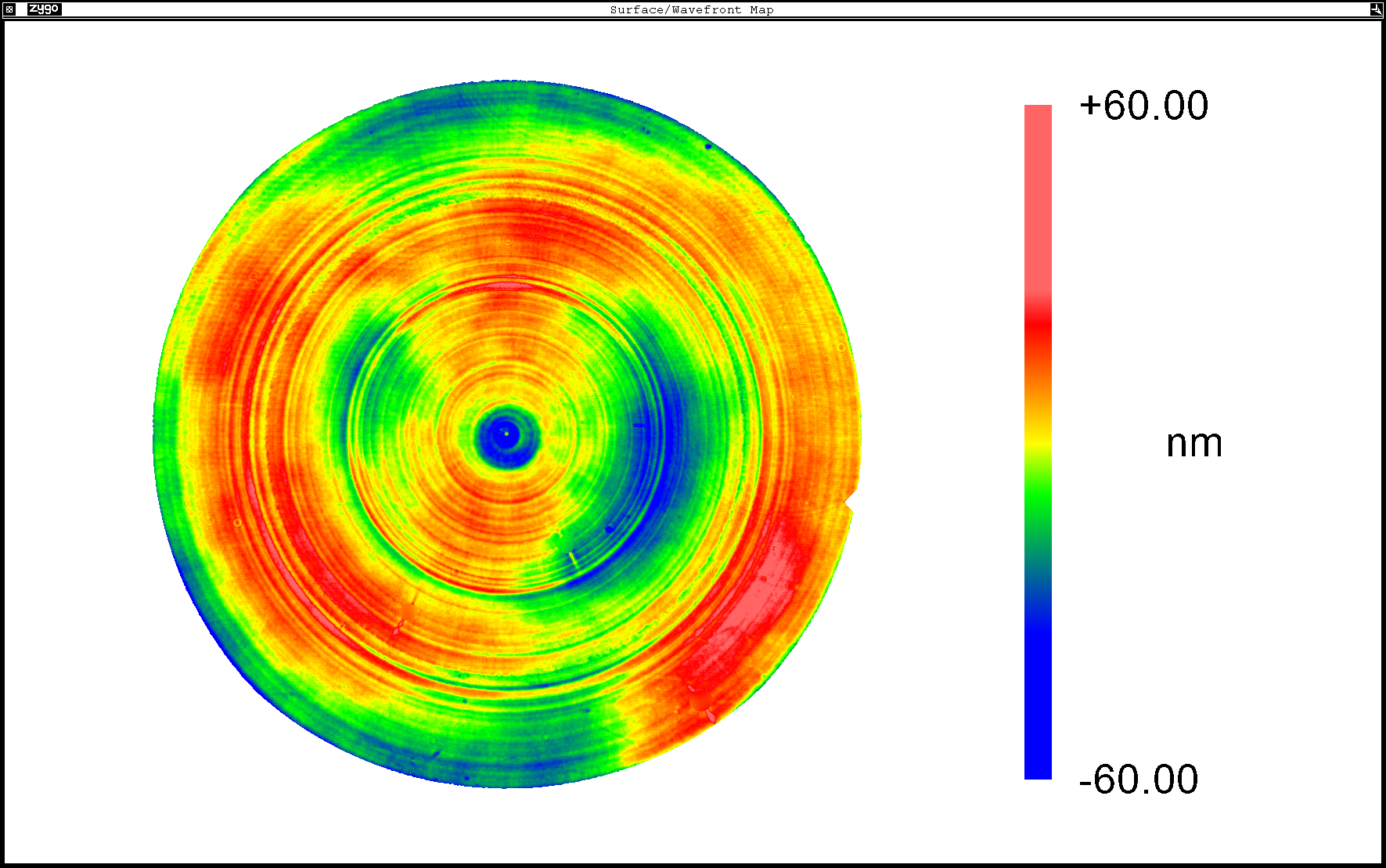}\\
			\begin{tabular}{cc}
				$p.-v.$ [nm] & $rms$ [nm] \\
				161.0 & 12.4 \\
			\end{tabular}
		\caption{Surface shape deviation after storage for two years at ambient conditions}
		\label{fig:twoyears}
	\end{minipage}
\end{figure}

In ground-based optical systems, particularly in space applications, the long-term
stability is an important cost factor. Metal mirrors can be designed to sustain a long
operating life. In order to achieve this, the mirror substrate material has to be
optimized regarding its dimensional stability. Even small changes in the material, e.g.,
because of residual stress, may cause the whole element to fail its function over time
\cite{Marschall:Maringer:1977}.
In order to evaluate the additive manufactured mirror regarding temporal stability,
the finished part was stored at ambient conditions for two years. In case of structural
modifications like creep or relaxation effects, the shape would change and by that
deteriorate the optical performance. Therefore, the interferometric shape measurement was
repeated after two years using the same measurement setup.
\Fref{fig:twoyears} shows the shape deviation after
this time duration. The measured mean values did not change significantly (\SI{12.5}{\nm}
rms) in comparison with the results shown in \fref{fig:FormCMP} (\SI{12.4}{\nm} rms), also
considering the measurement
accuracy of $\approx$~\SI{1}{\nm} rms (notice that the p.-v. values are not a good
indication of shape changes due to their sensitivity to outliers). By the process chain
applied, the material has been brought into a condition, where plastic deformations do not
occur at a temperature of $20~{}^\circ$C. Therefore, the mirror can be considered as
dimensionally stable at the ambient conditions taken into account.
In order to evaluate the influence of the mirror design on the shape accuracy
regarding the applied manufacturing chain, surface shape deviation is analyzed in more
detail. In contrast to classical mechanical designs the optical surface of the honeycomb
mirror is thinner (\SI{2}{\mm} thickness in CAD, \SI{>3}{\mm} in classical designs) and
could therefore be more prone to deformation under pressure (e.g. while polishing), where no
interior walls are present, while at supported areas the surface remains unchanged.
This would lead to an undesirable shape change, manifesting in a pattern as a
representation of the interior structure. 
\begin{figure}			
	%\centering
		\begin{minipage}{0.6\textwidth}
			\includegraphics[width=0.8\textwidth]{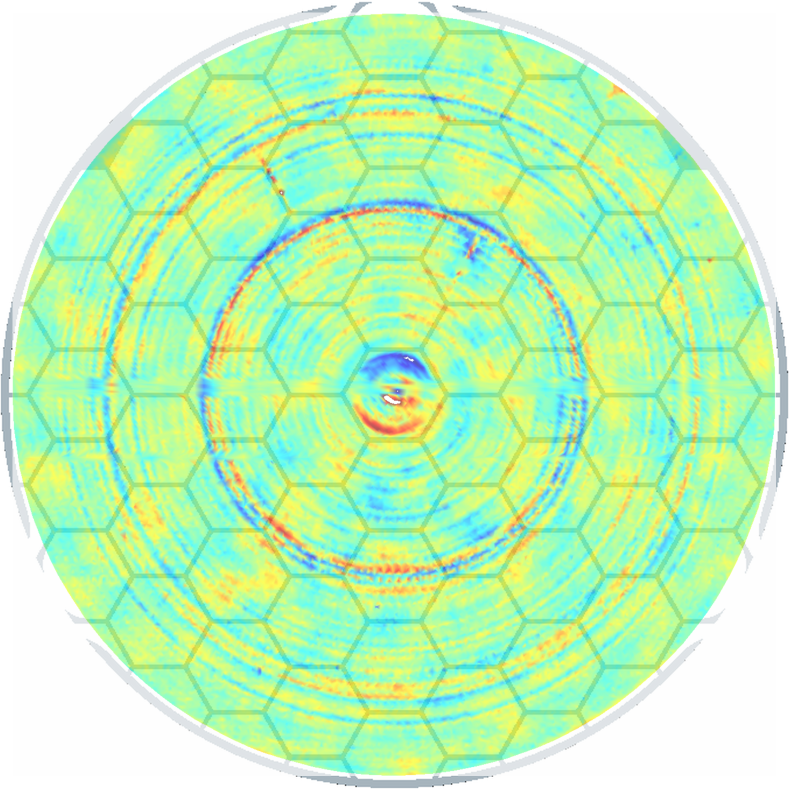}
				\label{fig:pt2}
		\end{minipage}
		%\hspace{0.05\textwidth}
		\begin{minipage}{0.2\textwidth}
			\includegraphics[height=5.5cm]{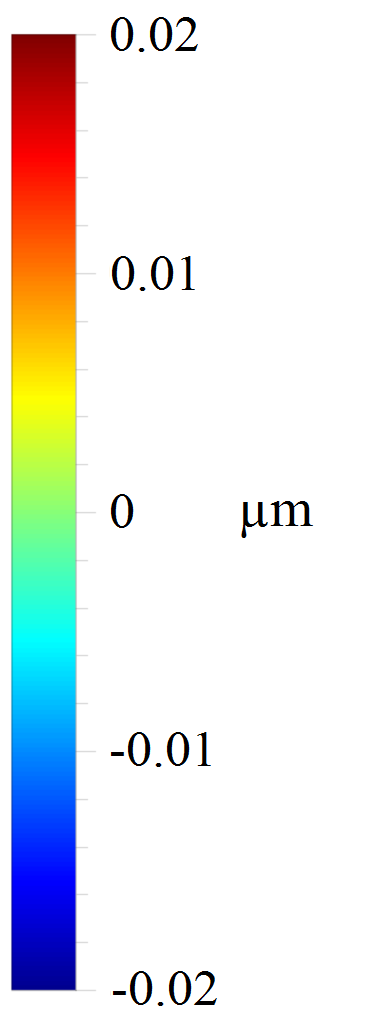}
		\end{minipage}
	\caption{Surface shape deviation with CAD interior overlay}
\end{figure}

The analysis was performed by
using the interferometric measurement (see \fref{fig:twoyears}) to get the deviation of
the mirror surface from the optical design pattern. This deviation contains several spatial
frequencies. The specific part of interest should have a frequency higher than the
standard surface form error. Therefore, all Zernike polynomials up to very high
orders were subtracted. This leads to visible small effects at the outer parts of the mirror
and repeating pattern of peaks that are up to \SI{7}{\nm} high and separated by a distance
of \SI{10}{\mm} which is almost equal to the hexagon diameter (see \fref{fig:pt2}). This
result shows that the mirror surface is stiff enough to resist manufacturing loads by
turning and polishing techniques, which is an indication that the design is well suited for
the applied manufacturing chain. It is expected, however, that a smaller diameter of the
hexagon cells is beneficial for stability.

% CONCLUSIONS AND OUTLOOK

\section{Conclusions and Outlook} \label{sec:outlook}
This study shows the suitability of Additive Manufacturing, i.e. Selective
Laser Melting, for the production of parts for precision applications. Aiming at an
improved mass reduction and high stiffness, a metal mirror design consisting of interior
honeycombs is developed. By exploiting the freedom of design from additive processes, it
is now possible to build such interior structures, while the backside remains closed, which
is beneficial for stiffness and simplifies the manufacturing. Numerical simulations show
that the honeycomb design outperforms conventional approaches.

The complex interior geometry
makes a quality assurance necessary, which X-ray tomography is a suitable tool to work
with. Missing or unintended structures can be visualized and
processes can be adapted, respectively. Due to the limited shape accuracy of SLM
fabrication, CAD models have to be tailored using offsets, where necessary.
The present work shows a complete manufacturing chain, including additive manufacturing,
diamond turning, magnetorheological finishing, and chemical mechanical polishing.
The shape accuracy and roughness, which were achieved, make the mirror substrate
suitable for optical applications up to the visible spectral range. 
As a result, it is shown that AlSi12 is a suitable aluminum alloy to generate metal mirrors,
in addition to AlSi10Mg and Al6061 which have been demonstrated in the literature so far \cite{Herzog:Segal:Smith:others:2015,Mici:Rothenberg:Brisson:others:2015}.
These findings are supported by temporal stability measurements under steady ambient
conditions which show no significant changes over a period of two years. 
Materials, which are thermally matched to electroless nickel (e.g. hypereutectic aluminum
silicon alloys) are of particular interest because of a reduced bimetallic bending during
varying temperatures \cite{Kinast:Hilpert:Lange:others:2014}. Aluminum with a silicon content of \SI{40}{wt\percent}, manufactured by SLM, is under present investigation.
The application of additive manufactured mirrors in space environment
requires further studies regarding the properties of the raw material.

\begin{figure}
	\centering
	\includegraphics[width=0.7\textwidth]{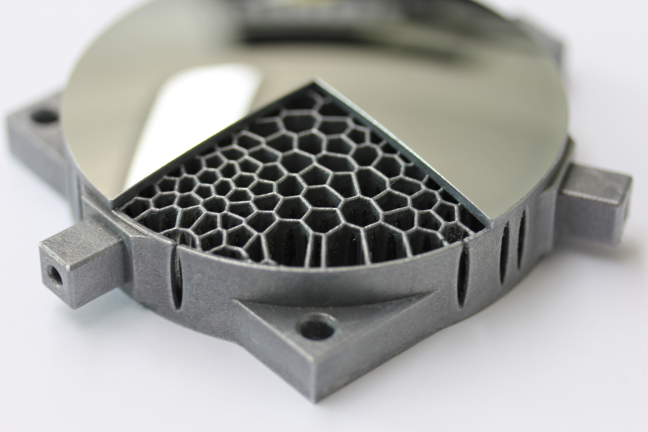}
	\caption{Diamond turned mirror with non-trivial interior structure}
	\label{fig:future}
\end{figure}

Future work will also address the development of load case optimized non-trivial designs
with a focus on even more lightweight mirrors.
\Fref{fig:future} shows an outlook on a possible advanced design. This demonstrator shows an
ultra precise diamond turned optical surface and a part of the interior structure, which was
manufactured using stochastic interior cells.
All features, like elliptical holes and smaller wall thickness were chosen due to the
results gathered within the present study. The design development as well as other
approaches will be the scope of a forthcoming publication.

\section*{Acknowledgements}
The authors are grateful to the involved colleagues at Fraunhofer~IOF, especially 
Robert~Jende, Matthias~Beier, Roland~Ramm, Christoph~Damm, and Nils~Heidler. 
They also want to thank Sebastian Scheiding and Marian Wiemuth for performing some
basic investigations and for many fruitful discussions.
Parts of the research were supported by the German Aerospace Center DLR within the 
project \mbox{\textit{ultraLEICHT}} under grant number 50EE1408.

\end{document}